\documentclass[10pt,letterpaper]{article}
\usepackage{opex3}
\usepackage{cite}

\begin{document}

%% NOTE: TITLE PAGE & TOC NOT USED FOR MANUSCRIPT SUBMISSIONS %%
%%\title{Simple template for authors submitting to \textit{Optics Express}}

\vskip4pc

%\tableofcontents
%\clearpage
%% NO TITLE PAGE FOR OPEX SUBMISSIONS %%

%% START HERE
%%%%%%%%%%%%%%%%%% title page information %%%%%%%%%%%%%%%%%%
%%%%%%%%%%%%%%%%% title page information %%%%%%%%%%%%%%%%%%
\title{Compact 2D nonlinear photonic crystal source of beamlike path entangled photons}

\author{E. Megidish,$^1$ A. Halevy,$^1$ H. S. Eisenberg,$^1$ A. Ganany-Padowicz,$^2$ N. Habshoosh,$^2$ and A. Arie$^2$}

\address{$^1$Racah Institute of Physics, Hebrew University of Jerusalem, \\ Jerusalem 91904, Israel\\ $^2$School of Electrical Engineering, Fleischman Faculty of Engineering, Tel Aviv University, \\ Tel Aviv 69978, Israel}
\email{hagaie@huji.ac.il} %% email address is required

%%%%%%%%%%%%%%%%%%% abstract and OCIS codes %%%%%%%%%%%%%%%%
%% [use \begin{abstract*}...\end{abstract*} if exempt from copyright]

\begin{abstract}
We demonstrate a method to generate entangled photons with
controlled spatial shape by parametric down conversion (PDC) in a
2D nonlinear crystal. A compact and novel crystal source was
designed and fabricated, generating directly path entangled
photons without the use of additional beam-splitters. This crystal
supports two PDC processes, emitting biphotons into two beamlike
modes simultaneously. Two coherent path entangled amplitudes of
biphotons were created and their interference observed. Our method
enables the generation of entangled photons with controlled
spatial, spectral and polarization properties.
\end{abstract}


%\ocis{(160.5298) Photonic crystals; (190.4410) Nonlinear optics,
%parametric processes; (270.0270) Quantum optics; (270.5585)
%Quantum information and processing.}

%%%%%%%%%%%%%%%%%%%%%%% References %%%%%%%%%%%%%%%%%%%%%%%%%

%%%%%%%%%%%%%%%%%%%%%%%%%%  body  %%%%%%%%%%%%%%%%%%%%%%%%%%
\section{Introduction}

Entangled photonic states are commonly generated using the
nonlinear process of optical parametric down-conversion (PDC)
\cite{Kwiat95}. In PDC, a single pump photon is converted into two
lower energy photons in a quadratic nonlinear crystal. As the pump
photon and the down-converted photons are of different
wavelengths, phase-matching (PM) conditions have to be fulfilled
in the form of momentum and energy conservation. As a result, the
down-converted photons are highly correlated in many degrees of
freedom. The PM conditions can be fulfilled in several ways, such
as angle and temperature tuning in birefringent crystals and with
the quasi phase matching (QPM) method, using periodically or
aperiodically poled crystals \cite{Armstrong62,Fejer}.

The QPM technique involves a spatial modulation of the
second-order nonlinearity of the material. As a result, the PM
criteria must be satisfied while taking into account the modulated
nonlinearity. The PM condition in $k$-space (the reciprocal
lattice) is therefore
\begin{equation}\label{PMinQPM}
\overline{k}_{2w}-\overline{k}_{w_1}-\overline{k}_{w_2}=\overline{G}_n\,,
\end{equation}
where $\overline{k}_{2\omega}$ is the momentum of the pump,
$\overline{k}_{\omega_i}$ the momentum of the down-converted
photons, and $\overline{G}_n$ is a reciprocal lattice vector of the
nonlinear crystal. The modulation vector and the material dispersion
properties determine which wavelengths satisfy the PM conditions.
Additionally, other properties of the generated beams can be
controlled by careful tuning of the modulation, such as their shape
\cite{Ellenbogen, Voloch12}, focusing \cite{Torres, Leng}, and
spectral properties \cite{Shiloh}. At first, QPM was realized by one
dimensional periodic modulation, but later more sophisticated
modulation patterns appeared, such as quasiperiodic modulation
\cite{Lifshitz, Zhu97} and two-dimensional periodic modulation
\cite{Berger98, Broderick00}. These patterns provide a larger
variety of reciprocal lattice vectors, hence offering more
possibilities for phase matching. When the nonlinear coefficient is
periodically modulated along two directions, a two-dimensional
nonlinear photonic crystal (NPC) is formed. This NPC supports
modulation vectors $\overline{G}_{m,n}=m G_x\hat{x}+n G_y\hat{y}$
with two degrees of freedom, such that PM can be fulfilled in
several directions and for different wavelengths.

Entanglement can be realized between various degrees of freedom of
the photons. In this work, we focus on entanglement between the
number of photons and their path. Such states are superpositions of
two $N$-photon amplitudes, in which all the photons are in one of
two possible optical modes or in the other. When the two modes are
interfered, oscillations that are $N$ times faster than the
wavelength are observed. These states have been shown to exceed the
classical limit of measurement accuracy and can be used to reach the
quantum Heisenberg limit \cite{Bollinger96}. Path entangled states
were demonstrated by Hong-Ou-Mandel (HOM) bunching photons at a
balanced beam-splitter (BS) \cite{Hong87, Rarity90}, combining a few
photons statistically with BSs \cite{Mitchel04}, post-selection of
appropriate amplitudes \cite{Nagata07}, and by combining classical
and nonclassical light beams at a BS \cite{Afek10}.

Ideally, the PDC process should generate the path entangled states
directly into easy to collect beamlike modes. However, in many
cases, owing to the rotational symmetry of the nonlinear process,
the down converted photons are generated along cones of light
\cite{Kwiat95}. This severely limits the collection efficiency of
the entangled photons. An alternative method is to generate the
entangled photons in two collinear counter-propagating directions
\cite{Kuklewicz, Kim06}. This scheme can provide high efficiency,
but requires pumping the nonlinear crystal simultaneously along two
counter-propagating directions, filtering of the collinear pump
beams, and an interferometric setup for combining the two
counter-propagating down-converted photons. Another alternative uses
two type-II crystals to generate non-collinear beamlike
polarization entangled photons \cite{Kim03}. None of these schemes
generates path entangled states directly.

\section{Theoretical background}
\begin{figure}[tb]
\centering\includegraphics[width=4in]{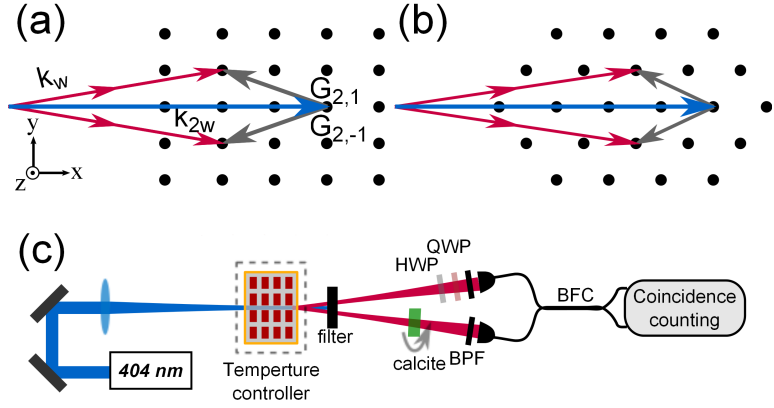}
\caption{\label{Fig1} (a) A reciprocal lattice
representation of the momenta involved in the down-conversion
processes in the crystal. (b) An alternative possible design that
avoids the $G_{2,0}$ circle. (c) The experimental setup (see main
text for details).}
\end{figure}

In this work, we demonstrate a path-entangled state that is
generated from a two-dimensional periodically poled NPC
\cite{Arie07}. This scheme, which was recently proposed
theoretically \cite{Gong12a}, overcomes the limitations of existing
methods by controlling the spatial properties of the down-converted
photons \cite{Torres}. The two biphoton states are directly
generated into well defined non-collinear beamlike modes, using a
single pump beam. The reciprocal lattice of the 2D NPC supports two
processes. Thus, a pump photon can split into two down-converted
photons that are in either one of two coherent well-defined spatial
modes [see Fig. \ref{Fig1}(a)]. The two-photon state is
\begin{equation}\label{State}
|\psi^\varphi\rangle=\frac{1}{\sqrt{2}}(|2,0\rangle+e^{2i\varphi}|0,2\rangle)\,,
\end{equation}
where $|a,b\rangle$ is a two-mode Fock state with $a$ photons in
one mode and $b$ in the other. The angle $\varphi$ is a
controllable relative phase between the two modes. The accumulated
phase between the two amplitudes is twice this angle due to the
two-photon effective de-Broglie wavelength \cite{Jacobson95}. The
control of the PM condition through the crystal temperature
affects the photons' mode shape and their overall collection
efficiency. The phase-matched process is for the $d_{zzz}$ element
of the quadratic nonlinear tensor, and thus the pump photon and
both down-converted photons are all extraordinarily polarized. The
reciprocal vector elements in our configuration are
\begin{eqnarray}\label{PM}
\nonumber &&mG_x=|\overline{k}_{2w}| -(|\overline{k}_{w_1}|+|\overline{k}_{w_1}|)\cos(\theta),\\
&&nG_y=(|\overline{k}_{w_1}|+|\overline{k}_{w_1}|)\sin(\theta).
\end{eqnarray}

\section{Crystal design and fabrication}
The NPC periods were calculated for a stoichiometric LiTaO$_3$ (SLT)
crystal \cite{IdoDolev}, where an angle of $0.8^\circ$ between the
pump and the down-converted photons inside the crystal was chosen.
For degenerated down-conversion process using $404\,$nm pump
photons, we found $\Lambda_x=3.2\,\mu$m and $\Lambda_y=13.46\,\mu$m
for m=n=1. In order to avoid a modulation period which is hard to
fabricate, we chose m=2, n=1 such that the $x$ period is
$\Lambda_x=6.4\,\mu$m.

For designing our poling pattern we assume a circular motif
\cite{Arie07} in each lattice point, i.e., the second order
nonlinear coefficient is inverted in circular areas that are
centered on each one of the lattice point. The radius R of the
circle determines the Fourier coefficient at each spatial frequency
and therefore the efficiency of the process. In order to avoid
overlap we limit the radius to be less than half of the period.
Since this is a rectangular lattice, according to \cite{Arie07} the
Fourier coefficient is given by
\begin{equation}\label{EqAdy}
{G}^{x}_{m,n}=\frac{2R}{\sqrt{(n\Lambda_x)^2+(m\Lambda_y)^2}}J_1[2\pi R\sqrt{(\frac{m}{\Lambda_x})^2+(\frac{n}{\Lambda_y})^2} ].
\end{equation}

The optimal result is obtained for a motif with a radius of $2.7\,
\mu m$, and the Fourier coefficient in this case is 0.087. A
$0.5\%$ Mg-doped SLT was poled by electric-field poling. The
crystal dimensions were $0.5\times13\times15\,$mm. The inverted
domain structure was created by applying an electrical field
through patterned electrodes on the crystal
surface\cite{Yamada92}. High-voltage pulses were applied to the
polar crystal surfaces using a computer-controlled system, by
limiting the pulse duration at a given current and driving voltage
level while monitoring the current and charge transfer. The total
switching charge required for complete poling was
60$\,\mu$C/cm$^{2}$, which was done with an electric field of
3\,KV/mm at room temperature. Under these conditions, the maximal
current density that we obtained was 25$\,\mu$A/cm$^2$. It should
be noted that when one uses small size of motif on the mask
(typically less than $5\,\mu$m), it is difficult to control the
exact shape and size of the inverted domains. To overcome this
limitation we use a multi-grating design, with different sizes and
shapes of the motif on the mask. The good quality of the realized
structure can be seen in the inset of Fig. \ref{Fig2}. Two
gratings have square motifs ($2.4\times2.4\,\mu$m and
$5.5\times5.5\,\mu$m), and two other gratings have rectangular
domains ($5.5\times6.73\,\mu$m and $2.4\times6.73\,\mu$m). The
inverted domains exhibited nearly circular shape when the square
motif was used, and an oval shape when the rectangular motif was
used. Each rectangular lattice was poled on an area of
~$3\times13\,$mm$^2$.

\section{Experimental setup and results}
A narrow line width diode laser with $35\,$mW and at a wavelength of
$404\,$nm pumps a two-dimensionally poled stoichiometric LiTaO$_3$
crystal whose temperature is controlled and stabilized within
$\pm0.05^\circ$C [see Fig. \ref{Fig1}(c)]. After passing through the
13\,mm long crystal, the pump light is filtered out using a dichroic
mirror and a long-pass filter. The $808\,$nm down-converted photons
are further filtered by a $3\,$nm wide bandpass filter (BPF) and
spatially filtered by coupling them into single mode fibers. A
half-wave plate (HWP) and a quarter-wave plate (QWP) before the
fiber coupling of one of the two paths are used to restore the
interference between the two paths, which is lost due to the random
polarization rotations in the two fibers. The relative phase between
the paths is controlled by rotating a $1\,$mm calcite in one of
them. After the two paths are combined at a balanced fiber coupler
(BFC), they are coupled into two single photon detectors. The two
paths coincidence count rate is detected within a $7\,$ns window
\cite{Beck}.

\begin{figure}[tb]
\centering\includegraphics[width=3.5in]{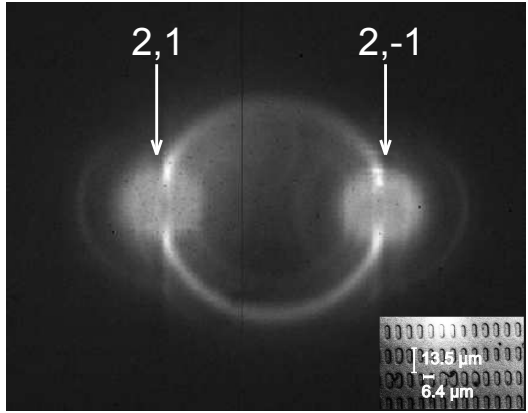}
\caption{\label{Fig2} A picture of the down-converted photons
through a 3\,nm bandpass filter centered at 808\,nm. It was taken
using a cooled sensitive camera placed after the pump filter [see
Fig. \ref{Fig1}(c)]. Inset: an optical microscope picture of the
NPC.}
\end{figure}

The angular spread of the down-converted photons was measured using
a cooled CCD camera placed after the pump block filter and the
narrow bandpass filter. The working condition was searched by tuning
the relative angle between the pump and the input facet, and the
crystal temperature. These angle and temperature values were set
such that the two down-converted modes were generated symmetrically
into beamlike modes. For our sample, such beam shape for the
$G_{2,1}$ and $G_{2,-1}$ modes was achieved at a crystal temperature
of $61.00\pm0.05^\circ$C, as presented in Fig. \ref{Fig2} by the two
filled circles. At lower temperatures, the difference between the
refractive indices of the pump and the down-converted photons is
smaller, the PM condition is not fulfilled, and no photons are
emitted into these modes. On the other hand, when the crystal
temperature is increased above this working point, the beamlike PM
pattern changes into cones, and its projection on the camera is in
the form of circles of increasing radius.

In order to observe the two-photon interference between the two
down-converted modes, first the coupling of each mode has to be
optimized. When blocking one mode before it is coupled into the
fiber, the coupling of the second mode into its respective fiber
is optimized through the two detectors' coincidence rate.
Similarly, the coupling of the first mode is optimized when the
second mode is blocked. Each of these two coincidence rates were
about 1000 per second for a pump power of $35\,$mW. The detection
of pairs of photons from each of the modes is not sufficient, as
an incoherent mixture of two photons in one of the two modes would
still show the same results. In order to demonstrate the presence
of coherence between the two process, as in the state of Eq.
\ref{State}, they are combined at the BFC. The ideal state of Eq.
\ref{State} is transformed by the BFC to
\begin{equation}\label{StateAfterBS}
|\psi'\rangle=\frac{1}{2}(1-e^{2i\varphi})|\psi^\frac{\pi}{2}\rangle+\frac{1}{2}(1+e^{2i\varphi})|1,1\rangle\,.
\end{equation}
When the relative phase between the two paths is scanned, the state
oscillates at double the phase between the state
$|\psi^\frac{\pi}{2}\rangle$ of both photons at a single mode, and
the state where their is a single photon in each of the modes. When
$\varphi=\pi/2$ or $3\pi/2$, the crystal generated output is an
eigenstate of the BFC operation. When $\varphi=0$ or $\pi$, the two
photons undergo a process which is the inverse of the HOM bunching
\cite{Hong87}, resulting in the $|1,1\rangle$ anti-bunched state.

\begin{figure}[tb]
\centering\includegraphics[width=3.5in]{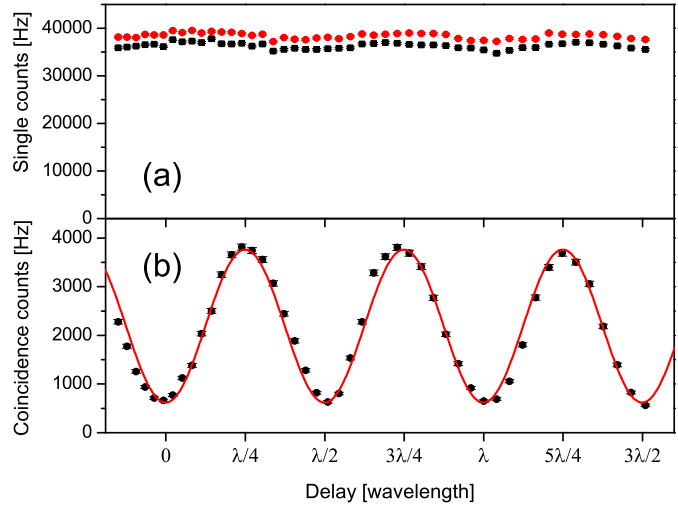}
\caption{\label{Fig3} (a) Single counts of the left (red circles)
or right (black squares) port of the BS. No dependence on the
relative delay between the two paths is observed. (b) Coincidence
counts as a function of the relative delay between the two paths.
Interference pattern in the coincidence counts is observed with a
contrast of $72\pm1\%$. The interference period is half of the
photons' wavelength. Error bars are calculated assuming Poissonian
noise statistics.}
\end{figure}

Clearly, an additional process is present in the form of a large
circle around the pump beam, overlapping with the two signal beams
(Fig. \ref{Fig2}). This unwanted process is a result of one
down-converted photon emitted along one beam while the other is
emitted along the other beam. The crystal PM vector for this
process is $G_{2,0}$. The number of events collected from this
unwanted process is measured as coincidence events between the two
modes. The ratio between the events produced by this unwanted
process and the desired $G_{2,\pm1}$ processes is $15\%$. Although
our results are not affected considerably by this process, we note
that a non-rectangular reciprocal lattice such as suggested in
Fig. \ref{Fig1}(b) will eliminate this process, as no reciprocal
vector lays at the middle point between the two participating
vectors. There are also three weak circles in the picture that we
attribute to other possible nonlinear processes.

We measured the single (one photon from either one of the BS
ports) and the coincidence counts (two photons exit the BS into
different ports) while changing one of the optical path lengths by
rotating the $1\,$mm calcite crystal (see Fig. \ref{Fig1}). The
single counts show no dependence to the delay changes whereas
quantum interference is observed in the coincidence counts
\cite{Edamatsu02}. Because two photons are accumulating phase
together, the total phase between the two terms of
$|\psi^\varphi\rangle$ is doubled. This coincidence corresponds to
the anti-bunched term of $|\psi'\rangle$. The observed
oscillations with a visibility of $72\pm1\%$ are presented in Fig.
\ref{Fig3}(b). The oscillation period is half the wavelength of
each of the photons, commonly regarded as super-resolution
\cite{Holland93}.

The non-perfect visibility is attributed to four causes. The fiber
coupler has a weak wavelength dependence, resulting in $45/55\%$
imbalance at the working wavelength. Secondly, there is inaccuracy
in the polarization matching of both modes at the BFC, resulting
in distinguishability between the two modes. Additionally, the
second order term, describing two pump photons that split into
four down-converted photons within the coherence time, reduces the
visibility. In order to evaluate the second order term effect, we
measured the number of pairs created in one path as a function of
the pump power. Deviation from a linear dependence indicates the
presence of higher order terms. We observed that $10\%$ of the
pairs for a 35\,mW pump power were created from high order events,
and estimate the total effect on the visibility to be a reduction
of $\sim7\%$. Lastly, the anti-bunched pairs from the $G_{2,0}$
circle affect the visibility. Ideally, they do not contribute to
the coincidence detection, regardless of the applied $\varphi$
phase \cite{Hong87}. Practically, the imbalance of the fiber
coupler can generate a constant background signal to the
coincidence interference. For the parameters of our experiment, we
calculated this background to reduce the visibility by 5\%.

\begin{figure}[tb]
\centering\includegraphics[width=3.5in]{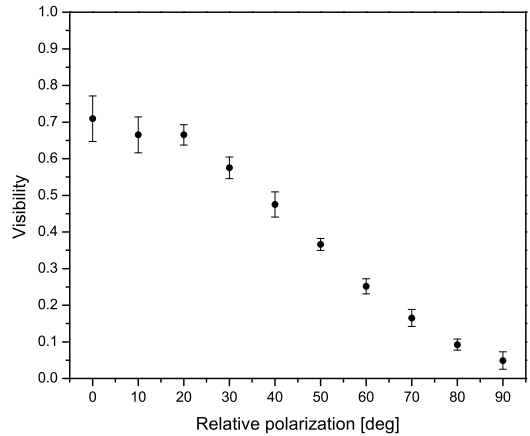}
\caption{\label{Fig4} The interference visibility as a function of
the relative polarization rotation between the two paths. At zero
angle the two paths are indistinguishable at the BFC and the
visibility is maximal. When the polarization of only one path is
rotated, distinguishability is introduced and the interference
contrast decreases.}
\end{figure}

The two-photon interference requires complete indistinguishability
between the photons in both paths. For the beam combining, spatial
overlap is achieved by using a single-mode BFC. Temporal
indistinguishability can be determined by the amount of spectral
filtering. In our experiment, the pump laser has a coherence time
of about $100\,$ns, which is much longer than the 300\,fs
coherence time imposed by the 3\,nm filters. Therefore, temporal
indistinguishability is not a problem. Polarization difference
between the two paths is another degree of freedom that can
contribute to distinguishability. As it is easily controlled using
the wave plates, it is possible to verify that the interference is
between the two spatial modes and not between two polarization
modes. We measured the interference visibility while gradually
rotating the photon's polarization on one path using a HWP (see
Fig. \ref{Fig4}). With no rotation, the polarizations of both
paths match, and interference is observed. When a relative
rotation between the two paths is applied, distinguishability is
introduced and the interference decreases. At $90^\circ$, there is
non-vanishing visibility of about 5\%, which is attributed to some
elliptical polarization mismatch. As stated above, this also
affects the maximal observed visibility.

%Another advantage for a cw laser pumping is the dispersion
%cancelation affect \cite{Steinberg} which allows us to change the
%optical path using a calcite crystal without creating temporal
%distinguishability.

\section{Conclusions}
In conclusion, we have demonstrated a method to control the spatial
properties of entangled down-converted photons. Two-photon path
entangled states were generated from a two-dimensional periodically
poled nonlinear crystal. Bunching on a beam splitter is not required
as the states are emitted directly from the source. The two paths of
the generated state were combined at a fiber coupler and quantum
amplitude oscillations with a doubled phase sensitivity were
observed. This source demonstrates the ability to simultaneously
phase-match more than one quantum process in such two-dimensional
crystals. This method can be further extended for generating
entangled photons with controlled spatial, spectral and polarization
properties. For example, polarization entangled photons can be
created using a slightly different scheme in which the same crystal
is pumped with $y$-polarized pump, thereby generating pairs of
orthogonally polarized photons through the $d_{yzy}$ process
\cite{Gong12b}.After the submission of this work, a similar independent result was submitted to the Arxiv \cite{Jin13}. 

\section*{Acknowledgments}
The authors thank the Israeli Ministry of Science for supporting
this work under Project 3-3445, and the Israel Science Foundation
for supporting this work under Grants 774/09 and 546/10.


\begin{thebibliography}{99}

\bibitem{Kwiat95} P. G. Kwiat, K. Mattle, H. Weinfurter, A. Zeilinger, A. V. Sergienko, and Y. Shih, ``New high intensity source of polarization-entangled photon pairs,'' \prl {\bf 75}, 4337--4341 (1995).
\bibitem{Armstrong62} J. A. Armstrong, N. Bloembergen, J. Ducuing, and P. S. Pershan, ``Interactions between light waves in a nonlinear dielectric,'' Phys. Rev. {\bf 127}, 1918--1939 (1962).

\bibitem{Fejer} M. M. Fejer, G. A. Magel, D. H. Jundt, and R. L. Byer, ``Quasi-phase-matched second harmonic generation: tuning and tolerances,'' IEEE J. Quant. Electron. {\bf 28}, 2631--2654 (1992).

\bibitem{Ellenbogen} T. Ellenbogen, N. Voloch-Bloch, A. Ganany-Padowicz, and A. Arie, ``Nonlinear generation and manipulation of Airy beams,'' Nat. Photon. {\bf 3}, 395--398  (2009).

\bibitem{Voloch12} N. Voloch-Bloch, K. Shemer, A. Shapira, R. Shiloh, I. Juwiler, and A. Arie, ``Twisting light by nonlinear photonic crystals,'' \prl {\bf 108}, 233902 (2012).

\bibitem{Torres} J. P. Torres, A. Alexandrescu, S. Carrasco, and L. Torner, ``Quasi-phase-matching engineering for spatial control of entangled two-photon states,'' \ol {\bf 29}, 376--378 (2004).

\bibitem{Leng} H. Y. Leng, X. Q. Yu, Y. X. Gong, P. Xu, Z. D. Xie, H. Jin, C. Zhang, and S. N. Zhu, ``On-chip steering of entangled photons in nonlinear photonic crystals,'' Nat. Commun. {\bf 2}, 429 (2011).

\bibitem{Shiloh} R. Shiloh and A. Arie, ``Spectral and temporal holograms with nonlinear optics,'' \ol {\bf 37}, 3591--3593 (2012).

\bibitem{Lifshitz} R. Lifshitz, A. Arie, and A. Bahabad, ``Photonic quasicrystals for nonlinear optical frequency conversion,'' \prl {\bf 95}, 133901 (2005).

\bibitem{Zhu97} S. N. Zhu, Y. Y Zhu, and N. B Ming, ``Quasi-phase-matched third-harmonic generation in a quasi-periodic optical superlattice,'' Science {\bf 278}, 843--846 (1997).

\bibitem{Berger98} V. Berger, ``Nonlinear photonic crystals,'' \prl {\bf 81}, 4136--4139 (1998).

\bibitem{Broderick00} N. Broderick, G. Ross, H. Offerhaus, D. Richardson, and D. Hanna, ``Hexagonally poled lithium niobate: a two-dimensional nonlinear photonic crystal,'' \prl {\bf 84}, 4345--4348 (2000).

\bibitem{Bollinger96} J. J. Bollinger, W. M. Itano, D. J. Wineland, and D. J. Heinzen, ``Optimal frequency measurements with maximally correlated states,'' \pra {\bf 54}, R4649--R4652 (1996).

\bibitem{Rarity90} J. G. Rarity, P. R. Tapster, E. Jakeman, T. Larchuk, R. A. Campos, and M. C. Teich, ``Two-photon interference in a Mach-Zehnder interferometer,'' \prl {\bf 65}, 1348--1351 (1990).

\bibitem{Hong87} C. K. Hong, Z. Y. Ou, and L. Mandel, ``Measurement of subpicosecond time intervals between two photons by interference,'' \prl \textbf{59}, 2044--2046 (1987).

\bibitem{Mitchel04}M. W. Mitchell, J. S. Lundeen, and A. M. Steinberg, ``Super-resolving phase measurements with a multiphoton entangled state,'' \nat {\bf 429}, 161--164 (2004).

\bibitem{Nagata07} T. Nagata, R. Okamoto, J. L. O'Brien, K. Sasaki, and S. Takeuchi, ``Beating the standard quantum limit with four-entangled photons,'' Science {\bf 316}, 726--729 (2007).

\bibitem{Afek10} I. Afek, O. Ambar, and Y. Silberberg, ``High-NOON states by mixing quantum and classical light,'' Science {\bf 328}, 879--881 (2010).

\bibitem{Kuklewicz} C. E. Kuklewicz, M. Fiorentino, G. Messin, F. N. C. Wong, and J. H. Shapiro, ``High-flux source of polarization entangled photons from a periodically poled KTiOPO$_4$ parametric down-converter,'' \pra {\bf 69}, 013807 (2004).

\bibitem{Kim06} T. Kim, M. Fiorentino, and F. N. C. Wong, ``Phase-stable source of polarization-entangled photons using a polarization Sagnac interferometer
,'' \pra {\bf 73}, 012316 (2006).

\bibitem{Kim03} Y.-H. Kim, ``Quantum interference with beamlike type-II spontaneous parametric down-conversion,'' \pra {\bf 68}, 013804 (2003).

\bibitem{Arie07} A. Arie, N. Habshoosh, and A. Bahabad, ``Quasi phase matching in two-dimensional nonlinear photonic crystals,'' Opt. Quant. Electron. {\bf 39}, 361--375 (2007).

\bibitem{Gong12a} Y.-X. Gong, P. Xu, Y. F. Bai, J. Yang, H. Y. Leng, Z. D. Xie, and S. N. Zhu, ``Multiphoton path-entanglement generation by concurrent parametric down-conversion
in a single $\chi^{(2)}$ nonlinear photonic crystal,'' \pra {\bf
86}, 023835 (2012).

\bibitem{Jacobson95} J. Jacobson, G. Bj\"{o}rk, I. Chuang, and Y. Yamamoto, ``Photonic de Broglie Waves,'' \prl {\bf 74}, 4835--4838 (1995).

\bibitem{IdoDolev} I. Dolev, A. Ganany-Padowicz, O. Gayer, A. Arie, J. Mangin, and G. Gadret, ``Linear and nonlinear optical properties of MgO:LiTaO$_3$,'' Appl. Phys. B {\bf 96}, 423--432 (2009).

\bibitem{Yamada92}  M. Yamada, N. Nada, M. Saitoh, and K. Watanabe, ``First-order quasi-phase matched LiNbO$_3$ waveguide periodically poled by applying an external field for efficient blue second-harmonic generation,'' Appl. Phys. Lett. {\bf 62}, 435--436 (1992).

\bibitem{Beck} D. Branning, S. Bhandari, and M. Beck, ``Low-cost coincidence-counting electronics for undergraduate quantum optics,'' Am. J. Phys. {\bf 77}, 667--670 (2009).

\bibitem{Edamatsu02} K. Edamatsu, R. Shimizu, and T. Itoh, ``Measurement of the photonic de broglie wavelength of entangled photon pairs generated by spontaneous parametric down-conversion,`` \prl \textbf{89},  213601 (2002).

\bibitem{Holland93} M. J. Holland and K. Burnett, ``Interferometric detection of optical phase shifts at the Heisenberg limit,'' \prl {\bf 71}, 1355--1358 (1993).

\bibitem{Gong12b} Y.-X. Gong, P. Xu, J. Shi, L. Chen, X. Q. Yu, P. Xue, and S. N. Zhu, ``Generation of polarization-entangled photon pairs via concurrent spontaneous parametric downconversions in a
single $\chi^{(2)}$ nonlinear photonic crystal,'' \ol {\bf 37},
4374--4376 (2012).

\bibitem{Jin13} H. Jin, P. Xu, X. W. Luo, H. Y. Leng, Y. X. Gong, and S. N. Zhu, "Compact engineering of path entangled sources from a monolithic quadratic nonlinear photonic crystal," http://arxiv.org/abs/1302.0162.
\end{thebibliography}
\end{document}